\documentclass[varenna,seceqno]{cimento}
\usepackage{graphicx,amssymb,amsthm,cite}

\title
{Comparison between quantum and classical dynamics \\ 
in the effective action formalism}
\author{Fabrizio Cametti,}
\institute{Dipartimento di Fisica, Universit\`a della Calabria, 
87036 Arcavacata di Rende, Italy 
and Istituto Nazionale per la Fisica della Materia, unit\`a di Cosenza}
\author{Giovanni Jona-Lasinio,}
\institute{Dipartimento di Fisica, Universit\`a di Roma ``La Sapienza'', 
Piazzale A. Moro 2, 00185 Rome, Italy}
\author{Carlo Presilla \atque}
\institute{Dipartimento di Fisica, Universit\`a di Roma ``La Sapienza'', 
Piazzale A. Moro 2, 00185 Rome, Italy
and Istituto Nazionale per la Fisica della Materia, 
unit\`a di Roma ``La Sapienza''}
\author{Fabio Toninelli}
\institute{Dipartimento di Fisica, Universit\`a di Roma ``La Sapienza'', 
Piazzale A. Moro 2, 00185 Rome, Italy}

\begin{document}
\maketitle

A major difficulty in comparing quantum and classical behavior resides
in the structural differences between the corresponding mathematical
languages. 
The Heisenberg equations of motion are operator equations only formally 
identical to the classical equations of motion.
By taking the expectation of these equations  
the well known Ehrenfest theorem provides identities 
which, however, are not a closed system of equations which allows to
evaluate the time evolution of the system.
The formalism of the effective action seems to offer a possibility 
of comparing quantum and classical evolutions in a systematic and 
logically consistent way by naturally providing approximation schemes
for the expectations of the coordinates which at the zeroth order 
coincide with the classical evolution \cite{jona94}.

The effective action formalism leads to equations of motion 
which differ from the classical equations 
by the addition of terms nonlocal in the time variable.
This means that for these equations an initial value problem is not
meaningful and they have to be interpreted in an appropriate way.
Here we analyze situations in which the nonlocal
terms can be reasonably approximated by local ones so that the quantum 
corrections do not modify the locality of classical equations. 
In the simplest approximation, the effective Lagrangian differs from 
the corresponding classical one by a renormalization of both the 
potential and the kinetic energy terms.
We shall not discuss the causal formalism used, for example, in Refs.
\cite{jordan,ch,cm}, as in the approximation considered this would
lead to the same local equations.

The present contribution describes the beginning of a systematic study of 
semiclassical evolutions using the effective action formalism.
In the first part, after introducing the formalism of the 
effective action and its expansion in powers of $\hbar$ (loop-expansion) 
in the context of quantum mechanics, we concentrate on the structure of 
the first order corrections in $\hbar$. 
These corrections are evaluated 
to the second order in the derivative expansion 
\cite{coleman-weinberg}, by two different methods. 
The first is based on a Euclidean approach \cite{IIM},
the second one on an adiabatic 
approximation in evaluating functional determinants.  

In the second part of the article we put the formalism at work, choosing 
as our case study a two-dimensional (2-D) anharmonic oscillator of 
the kind considered in molecular physics.
The results of the simulations show that by increasing $\hbar$ 
the effective dynamics tends to regularize the classical motion
and becomes qualitatively very similar to the quantum evolution
provided the energy is sufficiently small.

The evaluation of the effective dynamics in more general cases will
be presented in a forthcoming paper.

\section*{PART I}
\section{Effective action in quantum mechanics}
\label{azeff}

In this Section we define the effective action \cite{jona64}. 
For simplicity, consider a one degree of freedom Hamiltonian,
\begin{equation}
\hat H(\hat p,\hat q)=\hat H_0(\hat p,\hat q)+\hat U(\hat q),
\end{equation}
where
\begin{equation}
\hat H_0(\hat p,\hat q)=\frac {\hat p^2}{2m}+\frac 12 m\omega^2\hat q^2
\end{equation}
and the confining potential $\hat{U}(\hat{q})$ is an even polynomial of 
$\hat{q}$.
We choose the constant of $\hat{U}(\hat{q})$ so that the lowest eigenvalue
of $\hat H$ is 0.
The generating functional of the Green functions is
\begin{eqnarray}
\label{z1}
&&Z[J]=\langle0|{\cal T}(e^{\frac{i}{\hbar}\int dt
J(t)\hat{q}(t)})|0\rangle
\end{eqnarray}
where $| 0\rangle$ is the ground state of $\hat H$, $\hat{q}(t)=
e^{\frac i{\hbar} \hat Ht}\,\hat q\,e^{-\frac i{\hbar} \hat Ht}$, $J(t)$ is a source vanishing for $|t|
\rightarrow\infty$ and ${\cal T}$ is the time-ordering operator.
In Eq. (\ref{z1}), as well as in the following, the integrations 
with boundaries not explicitly indicated are to be understood between 
$-\infty$ and $+\infty$.
The generating functional of the connected Green functions is defined as
$W[J]=-i\hbar\ln Z[J]$ and the Legendre transform of $W[J]$ gives the effective 
action. By indicating with $q$ the variable conjugated to $J$, 
\idest, 
\begin{equation}
q(t)=\frac{\delta W[J]}{\delta J(t)},
\label{qc}
\end{equation}
we define
\begin{equation}\label{gamma}
\Gamma[q]=W[J]-\int dt q(t)J(t),
\label{eadef}
\end{equation}
where $J$ has to be thought, inverting relation (\ref{qc}),
as a functional of $q$.
The functional $\Gamma[q]$ represents the analog of the classical 
action, $S[q]=\int dt \left( \frac{1}{2} m \dot{q}^2(t) -V(q(t)) \right)$,
where $V(q)=\frac{1}{2} m \omega^2 q^2 +U(q)$,
and can be written in the form
\begin{equation}
\Gamma[q]=S[q]+\tilde\Gamma_\hbar[q],
\end{equation}
with $\tilde\Gamma_0[q]=0$.
The Legendre transform can be calculated using the methods of 
Ref. \cite{DJL}.

From the functional derivative of the classical action 
with respect to the position $q(t)$ one obtains the
Euler-Lagrange equation of motion
\begin{equation}\label{classica}
\frac{\delta S[q]}{\delta q(t)}=-J(t).
\end{equation}
In the same way the functional derivative of the effective action 
$\Gamma[q]$ with respect to the $q(t)$ given by (\ref{qc}) yields 
\begin{equation}
\frac{\delta \Gamma[q]}{\delta q(t)}=-J(t).
\label{eulaq}
\end{equation}
This equation can be rewritten in the form
\begin{eqnarray}
\label{label}
m \ddot q(t)+\partial_q V(q(t))-\frac{\delta\tilde\Gamma_\hbar[q]}{\delta
q(t)}=J(t).
\end{eqnarray}
As we shall see in the next Section, $\tilde\Gamma_\hbar[q]$ admits
an expansion in powers of $\hbar$, whose coefficients have a simple 
diagrammatic interpretation (loop-expansion). In this way we can view
the quantum integro-differential equation (\ref{label}) as a perturbation
of the classical equation of motion.

In order to interpret the solutions of Eq. (\ref{eulaq}), we rewrite 
$Z[J]$, defined in (\ref{z1}), in the equivalent form
\begin{equation}
\label{equiva}
Z[J]=\langle0|U^{J}_{S}(+\infty,-\infty)|0\rangle=
\langle0|{\cal T}(e^{-\frac i{\hbar}\int dt 
[\hat H-\hat qJ(t)]})|0\rangle,
\end{equation}
where $U^J_S(t_b,t_a)$ is the evolution operator 
from $t_a$ to $t_b$ in the Schr\"odinger representation and in presence 
of the external source $J(t)$. 
Note that in quantum mechanics $J(t)$ is an external force.
The variable conjugated to $J$ then is
\begin{equation}
\label{qqw2}
q(t)=\frac{\langle\alpha_{[J]}|U^{J}_S(t,0)^\dag\,
\hat q\,U^J_S(t,0)|\beta_{[J]}\rangle}{\langle\alpha_{[J]}|
\beta_{[J]}\rangle},
\end{equation}
where
\begin{eqnarray}
\label{alfabeta2}
&&\langle\alpha_{[J]}|=\langle0|U^{J}_{S}(+\infty,0)
\\ 
&&|\beta_{[J]}\rangle=U^{J}_{S}(0,-\infty)|0\rangle.
\end{eqnarray}
Since generally $|\beta_{[J]}\rangle$  differs from 
$|\alpha_{[J]}\rangle$, $q(t)$ 
is a nondiagonal matrix element of $\hat q$ between two states which
evolve in presence of $J(t)$. 
The solution of Eq. (\ref{eulaq}) therefore can be complex valued.
In the harmonic case $U(q)$ constant, $|\alpha_{[J]}\rangle$ and 
$|\beta_{[J]}\rangle$ are coherent states and coincide, up to a phase,
if $\tilde J(\omega)=0$, where $\tilde J$ is the Fourier transform of the external force. In the anharmonic case, more complicated 
conditions have to be imposed on $J$ so that the two states coincide.
If these conditions are satisfied, $q(t)$ is the expectation value 
of the position operator.

\section{Loop expansion of the effective action}
\label{par:semicl}

The effective action cannot be evaluated exactly for anharmonic
systems, \idest, $U(q)\neq$ constant. A widely used approximation scheme 
is the loop expansion (see for example \cite{Jackiw,AMIT}), 
or semiclassical approximation, consisting in an expansion of 
$\Gamma[q]$ in powers of $\hbar$. 
At the lowest order the effective action coincides with the
classical action, whereas the one-loop term is expressed by means of 
a functional determinant.

In order to obtain the loop expansion we express $Z[J]$ as a path 
integral.
Equation (\ref{equiva}) can be rewritten, using the Gell-Mann and
Low theorem \cite{fetter}, as
\begin{equation}
\label{gelo}
Z[J]=\frac{\langle0_0|{\cal T}(e^{-\frac i{\hbar} \int dt [\hat U(\hat q_0(t))-
\hat q_0(t)J(t)]})|0_0\rangle}
{\langle0_0|{\cal T}(e^{-\frac i{\hbar} \int dt \hat U(\hat q_0(t))})|0_0\rangle},
\end{equation}
where $\hat q_0(t)=e^{\frac i{\hbar} \hat H_0t}\,\hat q\,e^{-\frac i{\hbar} \hat H_0t}$ 
and $|0_0\rangle$ is the ground state of $\hat H_0$.
Equation (\ref{gelo}) is equivalent to
\begin{equation}\label{zj}
Z[J]=\frac{\left.e^{-\frac i{\hbar} \int dt U\left(\frac \delta
{\delta J'(t)}\right)}Z_0[J']\right|_{J'=J}}
{\left.e^{-\frac i{\hbar} \int dt U\left(\frac \delta
{\delta J'(t)}\right)}Z_0[J']\right|_{J'=0}},
\end{equation}
where $Z_0[J]$ in terms of Feynman path integrals \cite{feynhibbs} reads
\begin{equation}
\label{z0}                                 
Z_0[J]=\lim_{T\rightarrow\infty}\int dx\,dy \int d[q]^y_x\,
e^{\frac i{\hbar}\int_{-T}^T dt \left[\frac m2\dot q ^2(t)-\frac m2\omega^2q^2(t)+
J(t)q(t)\right]}\varphi_0(x)\varphi_0(y).
\end{equation}
Here $\varphi_0(x)=\langle x|0_0\rangle$ and $d[q]^y_x$ is the functional measure
on paths with endpoints $q(-T)=x,\,q(T)=y$. 
The purely oscillating integrand in Eq. (\ref{z0})
can be regularized by changing $\omega$ into 
$\omega_\varepsilon\equiv\omega(1-i\varepsilon)$ with 
$\varepsilon\rightarrow0^+$ \cite{ramond}.
Comparing Eqs. (\ref{zj}) and (\ref{z0}) we obtain 
\begin{eqnarray}
\label{teo2}
Z[J]&=&\!\lim_{\varepsilon\rightarrow 0^+}\!\lim_{T\rightarrow\infty}
\\ &&
\frac{\int dx\,dy \int d[q]^y_x\,
e^{\frac i{\hbar} \int_{-T}^T dt \left[\frac m2\dot q ^2(t)-\frac m2\omega_\varepsilon^2q^2(t)
-U(q(t))+J(t)q(t)\right]}\varphi_0(x)\varphi_0(y)}
{\int dx\,dy \int d[q]^y_x\,
e^{\frac i{\hbar} \int_{-T}^T dt \left[\frac m2\dot q ^2(t)-\frac m2 \omega_\varepsilon^2q^2(t)
-U(q(t))\right] }\varphi_0(x)\varphi_0(y)}.
\nonumber
\end{eqnarray}

Now we apply the stationary phase approximation to (\ref{teo2}), 
expanding the exponent 
at the numerator around the solution $q_0(t)$ of
\begin{equation}
m\ddot q_0(t)=-m\omega^2_\varepsilon q_0(t)-\partial_qU(q_0(t))+J(t)
\end{equation}
which vanishes for $|t|\rightarrow\infty$. We find
\begin{eqnarray}
\label{gauss}
Z[J]&\simeq&e^{\frac i{\hbar}\left(S[q_0]+\int dt J(t)q_0(t) 
\right)}\\
&&\times\lim_{\varepsilon\rightarrow 0^+}\lim_{T\rightarrow\infty}\frac{\int d[q]^0_0\,e^{\frac i{2\hbar}
\int_{-T}^T dt \left[\dot q^2(t)-\omega_\varepsilon^2q^2(t)-\frac1m\partial_q^2U(q_0(t))q^2(t)\right] }}
{\int d[q]^0_0\,e^{\frac i{2\hbar}\int_{-T}^T dt \left[\dot q^2(t)-
\omega_\varepsilon^2q^2(t)\right] }}.
\nonumber
\end{eqnarray}
Note that the integrations over $x$ and $y$ disappear since 
$\varphi_0(.)$ is
proportional to $\delta(.)$ in the limit $\hbar\to 0$.
The Gaussian integrals in (\ref{gauss}) can be performed yielding
\begin{eqnarray}
\label{zej}
Z[J]&\simeq&e^{\frac i{\hbar}\left(S[q_0]+\int dt J(t)q_0(t) \right)}\\
&&\times\lim_{\varepsilon\rightarrow 0^+}\lim_{T\rightarrow\infty}\left(\frac{\det\left(-\partial^2_t
-\omega_\varepsilon^2-\frac1m\partial_q^2U(q_0(t))\right)}{\det\left(-\partial_t^2-
\omega_\varepsilon^2\right)}\right)^{-\frac 1 2}
\nonumber
\end{eqnarray}
where the differential operators act on functions $y(t)$ with Dirichlet
boundary conditions $y(-T)=y(T)=0$.
From Eq. (\ref{zej}) we obtain
\begin{eqnarray}
W[J]&=&W_0[J]+\hbar\,W_1[J]+{\cal O}\,(\hbar^2) \\
&=&S[q_0]+\int dt J(t)q_0(t) \nonumber\\ 
&&+\lim_{\varepsilon\rightarrow0^+}\lim_{T\rightarrow\infty}
\frac{i\hbar}2\ln\left(\frac{\det\left(-\partial_t^2
-\omega_\varepsilon^2-\frac1m\partial_q^2U(q_0(t))\right)}{\det\left(-\partial_t^2-
\omega_\varepsilon^2\right)}\right)
+{\cal O}\,(\hbar^2).
\nonumber
\end{eqnarray}
Setting $q=q_0+\hbar q'$ and remembering that
$\left.\frac{\delta S[q]}{\delta q(t)}\right|_{q_0}=-J(t)$, the
effective action to one-loop order is
\begin{eqnarray}
\label{aa}
\Gamma[q]&=&\Gamma_0[q]+\hbar\,\Gamma_1[q]+{\cal O}\,(\hbar^2) \\
&=&W_0[J]+\hbar W_1[J]-
\int dt q(t)J(t) +{\cal O}\,(\hbar^2)\nonumber\\ 
&=&S[q-\hbar q']+\hbar W_1[J]- \hbar \int dt q'(t)J(t) +{\cal O}\,(\hbar^2)\nonumber\\ 
&=&S[q]+\frac{i\hbar} 2 \lim_{\varepsilon\rightarrow 0^+}
\lim_{T\rightarrow\infty}\left.\ln\left(\frac{\det\left(-\partial_t^2
-\omega_\varepsilon^2-\frac1m\partial_q^2U(q(t))\right)}{\det\left(-\partial_t^2-
\omega_\varepsilon^2\right)}\right)\right|_
{\mbox{\tiny{ Dirichlet($\pm T$)}}}\hspace{-1.5cm}+{\cal O}\,(\hbar^2).
\nonumber
\end{eqnarray}

\section{Derivative expansion of the effective action}
\label{derivative}

The classical action $S[q]$ is the time integral of a density
(the Lagrangian) which is an ordinary function of
$q(t)$ and $\dot q(t)$. 
As a consequence, the classical equation of motion (\ref{classica}) 
is a differential equation.
On the other hand, the effective action $\Gamma[q]$ is nonlocal in
time and, therefore, the variational equation (\ref{eulaq}) is also
nonlocal. 
If $q(t)$ varies slowly, however, it is possible to expand $\Gamma[q]$
around a constant value of $q$ (derivative expansion
\cite{IIM,miranski}). 
In this expansion one finds that also
$\Gamma[q]$ can be written as the time integral of a density, which is a
series of terms involving time derivatives of $q(t)$ of increasing order:
\begin{eqnarray}
\label{qloc}
\Gamma[q]=\int dt \left(-V_e(q(t))+\frac{Z(q(t))}2\dot q^2(t)+A(q(t))\dot q^4(t)+
B(q(t))\ddot q^2(t)+\ldots\right).
\end{eqnarray}
As we shall see, the derivative expansion (\ref{qloc}) does
not generally converge and has only an asymptotic validity for 
$q(t)\rightarrow$ constant.
The absence in (\ref{qloc}) of odd powers of $\dot q(t)$
is a consequence of the time reversal symmetry of the Hamiltonian.

Except for $V_e$ and $Z$, all the terms in the derivative expansion (\ref{qloc}) are at least of order $\hbar$:
\begin{eqnarray}
\label{qpote}
&&V_e(q)=\frac 12 m q^2\omega^2+U(q)+\hbar V_{e1}(q)+
{\cal O}\,(\hbar^2)\\ 
\label{qzeta}
&&Z(q)=m+\hbar\, Z_1(q)+{\cal O}\,(\hbar^2)
\\ 
&&A(q)=\hbar\, A_1(q)+{\cal O}\,(\hbar^2)
\\
&&B(q)=\hbar\, B_1(q)+{\cal O}\,(\hbar^2).
\end{eqnarray}
The effective
potential $V_e(q)$, well known in quantum field theory in the study of
spontaneous symmetry breaking \cite{coleman-weinberg}, is everywhere 
convex \cite{fukuda}. 
It may happen that the effective potential evaluated at a finite 
$\hbar$ order loses somewhere its convexity 
if the classical potential is not everywhere convex \cite{fujimoto}.
In this paper we restrict ourselves to a phase-space region where 
the evaluated effective potential is convex.

If the derivative expansion (\ref{qloc}) is truncated at a finite 
order $2N$, the corresponding variational equation is a differential 
equation of order $2N$. 
We thus have a Cauchy problem with $2N$ initial conditions. 
It is clear that these conditions do not determine completely
the initial wave function of the system.
They are constraints which must be imposed in the choice of
the initial wave function for a comparison between true and effective
quantum evolutions.
We confine ourselves to the 
second order in the derivative expansion (DE2), that is 
\begin{equation}
\label{qlocv}
\Gamma[q]\simeq\int dt \left(-V_e(q(t))+\frac{Z(q(t))}2\dot q^2(t)\right).
\end{equation}
This is the simplest approximation to the effective action which 
preserves the structure of the classical equations of motion.

In the following we work out and compare two methods to obtain the
derivative expansion of the effective action.
The first is an adaptation to quantum mechanics of a method \cite{IIM}
used in quantum field theory and based on the Euclidean 
functional formalism. 
In the second method, we relate the derivative expansion
to the adiabatic approximation of a differential equation 
with slowly varying coefficients. 
In this way we are able to give an estimate of the validity
of the derivative expansion.

\subsection{Derivative expansion: Euclidean approach}
\label{aiut!}

The derivative expansion of the effective action can be obtained 
starting from the Euclidean generating functional
\begin{equation}
\label{last?}
Z_E[J]= \frac{\int d[q]^y_x\,
e^{-\frac {1}{\hbar}\left(S_E[q]-\int dt J(t)q(t)\right) }
\varphi_0(x)\varphi_0(y)\,dx\,dy}
{\int d[q]^y_x\,e^{-\frac{1}{\hbar}S_E[q]}\varphi_0(x)\varphi_0(y)\,dx\,dy},
\end{equation}
where the Euclidean action $S_E[q]$ is defined by
\begin{equation}
\label{azcleu}
S_E[q]=\int dt \left(\frac12 m\dot q^2(t)+\frac12 m\omega^2q^2(t)+U(q(t))\right).
\end{equation}
Setting $W_E[J]=\hbar \ln Z_E[J]$ and $q(t)=
\frac{\delta W_E[J]}{\delta J(t)}$, we introduce the Euclidean effective action
\begin{equation}
\Gamma_E[q]= W_E[J]-\int dt J(t)q(t).
\end{equation}
In analogy with the results of Section \ref{par:semicl}, 
to one-loop order we have
\begin{equation}
\label{ge}
\Gamma_E[q]=-S_E[q]-\frac\hbar2 \ln\frac{\det\left[\frac{\delta^2 S_E[q]}{\delta q(t)\delta q(s)}\right]}
{\det\left[\frac{\delta^2 S_E[0]}{\delta q(t)\delta q(s)}\right]}+{\cal O}(\hbar^2),
\end{equation}
where the differential operator
$\frac{\delta^2 S_E[q]}{\delta q(t)\delta q(s)}$ can be rewritten as
\begin{equation}
\frac{\delta^2 S_E[q]}{\delta q(t)\delta q(s)} =
\left[-m\partial_t^2+m\omega^2+\partial^2_qU(q(t))\right]\delta(t-s).
\end{equation}
The second order in the derivative expansion of the Euclidean 
effective action is
\begin{equation}
\label{ge2}
\Gamma_E[q]=-\int dt\left[V_e(q(t))+\frac12 Z(q(t)) \dot q^2(t)\right],
\end{equation}
where $V_e(q)$ and $Z(q)$ are the same functions that appear in (\ref{qloc}).

The effective potential can be found by combining (\ref{ge}) and 
(\ref{ge2}) for $q(t)$ constant
\begin{equation}
\label{logmia}
\int dt V_{e1}(q)=\frac12\ln \det\left[\frac{\delta^2 S_E[q]}{\delta q(t)\delta q(s)}\right]
-\frac12\ln \det\left[\frac{\delta^2 S_E[0]}{\delta q(t)\delta q(s)}\right].
\end{equation}
Employing the functional analogue of  the identity
$\ln \det {\mathbb{A}}= \mbox{tr} \ln {\mathbb A}$, valid for any Hermitian matrix ${\mathbb A}$, we get
\begin{equation}
\ln \det\left[
\frac{\delta^2 S_E[q]}{\delta q(t)\delta q(s)}\right]=
\mbox{tr} \ln\left[\frac{\delta^2 S_E[q]}{\delta q(t)\delta q(s)}\right].
\end{equation}
We use the Dirac notation to write
\begin{equation}
\left(-\partial_t^2+ \omega^2+\frac1m\partial^2_qU(q)\right)\delta(t-s)
= \langle t|\left(\hat{P}^2+ \omega^2+\frac1m\partial^2_qU(q)\right)|s\rangle,
\end{equation} 
where the $\hat{P}$ operator is defined in the $\{|t\rangle\}$ basis by 
$\langle t|\hat{P}|s\rangle=-i\frac{\partial}{\partial t}\delta(t-s)$.
Equation (\ref{logmia}) becomes
\begin{equation}
\label{uffa}
~~~~~~~~~
\int dt V_{e1}(q)=\frac12 \int dt\left\{\langle t|\ln\left(\hat{P}^2+
\omega^2+\frac1m\partial^2_qU(q)\right)|t\rangle-
\langle t|\ln\left(\hat{P}^2+\omega^2\right)|t\rangle \right\}.
\end{equation}
With the help of the identity $\int dp |p\rangle \langle p|=1$, where 
$\hat{P} |p\rangle = p |p \rangle$, we can write
\begin{equation}
\label{labelmia}
\langle t|\ln\left(\hat{P}^2+\omega^2+\frac1m\partial^2_qU(q)\right)|t\rangle=
\frac1{2\pi}\int dp\ln\left(p^2+\omega^2+\frac1m\partial^2_qU(q)\right).
\end{equation}
The integral in the above expression can be evaluated exactly
and from (\ref{uffa}) we finally get
\begin{equation}
V_{e1}(q)=\frac12  \left(\sqrt{\omega^2+\frac1m\partial^2_qU(q)}-\omega\right). 
\end{equation} 

The determination of $Z_1(q)$ is more involved.
From Eq. (\ref{ge2}) we see that $Z(q)$ is the coefficient of the term 
containing $\dot q(t)^2$ in the effective action.
We can thus write
\begin{equation}
\int dt Z_1(q(t)) \dot{q}^2(t)=\ln \det\left[
\frac{\delta^2 S_E[q] }{\delta
q(t)\delta q(s)} \right]-
\left( \ln\det\left[ \frac{\delta^2 S_E[q_c]}{\delta
q(t)\delta q(s)} \right] \right)_{q_c \to q(t)},
\label{bohmia}
\end{equation}
with the assumption that we  consider in the r.h.s. only those 
terms with at most two time derivatives of $q(t)$. 
The first term in the r.h.s. of (\ref{bohmia}) is essentially the 
one-loop 
Euclidean effective action, while the second one comes from the effective 
potential.
The second functional determinant in (\ref{bohmia}) has to be 
evaluated with a constant $q_c$ which, at the end,  
must be replaced with $q(t)$. 
The terms due to the normalization of $Z_E[J]$, being common to both the 
effective action and the effective potential, cancel each other. 
Again we change the logarithm of the determinant
into the trace of the logarithm and write the differential operators 
in Dirac notation. 
It is useful to introduce two operators, $\hat{P}$ and $\hat{T}$, 
satisfying the commutation relation $[\hat{T},\hat{P}]=i$, 
and with elements 
$\langle t|\hat{P}|s\rangle=-i\frac{\partial}{\partial t}\delta(t-s)$,
$\langle p|\hat{T}|q\rangle=i\frac{\partial}{\partial p}\delta(p-q)$,
where $|t\rangle$,$|s\rangle$ and $|p\rangle$, $|q\rangle$ are 
eigenstates of $\hat{T}$ and $\hat{P}$, respectively. 
In addition, we write the difference of two logarithms of positive 
defined operators in the parametric form
\begin{equation}
\label{parametrica}
\ln \hat{A}-\ln \hat{B}=\int_0^\infty\frac{ds}{s}
\left(e^{-\hat{B}s}-e^{-\hat{A}s} \right).
\end{equation}
We then arrive at the following expression
\begin{eqnarray}
\label{basta!}
&&\int dt Z_1( q(t)) \dot{q}^2(t) \\ 
&&=\int dt\langle t|\int_0^\infty\frac{ds}{s}
e^{-(m\hat {P}^2+m\omega^2+\partial^2_qU(q(t)))s} 
-e^{-(m\hat {P}^2+m\omega^2+\partial^2_qU(q(\hat T)))s}|t\rangle.
\nonumber
\end{eqnarray}
Since we keep only terms at most quadratic in
$\dot q(t)$, we can expand $\partial^2_qU(q(\hat{T}))$ as follows:
\begin{equation} 
\partial^2_qU(q(\hat{T}))=\partial^2_qU(q(t))+\hat Q a(t) +\frac12 \hat{Q}^2b(t),
\end{equation}
where $\hat Q=\hat T-t$, $a(t)=\partial_t\partial^2_qU(q(t))$ and $b(t)=\partial^2_t
\partial^2_qU(q(t))$. All the terms proportional to $\hat{Q}^n$, with $n\geq3$, are neglected since they do not contribute to the determination of 
$Z_1(q)$.
The expressions for $\langle t| e^{-m\hat{P}^2s}|t\rangle $ and for
$\langle t| e^{-\left[ m\hat{P}^2+\hat Qa(t)+\frac12 \hat{Q}^2 b(t)\right]s}|t\rangle $ are known
\cite{schulman} and can be inserted in (\ref{basta!}). 
Finally we can expand the
integrand in Eq. (\ref{basta!}) maintaining only the terms linear in $b(t)$ and at most
quadratic in $a(t)$. Performing the integration over the variable $s$, 
we obtain
\begin{equation}
Z_1(q)=\frac 1 {32m^2}\frac{(\partial_q^3U(q))^2}{\left(\omega^2+\frac1m\partial_q^2U(q)
\right)^\frac 52}.
\end{equation}

\subsection{Derivative expansion as a WKB-like approximation}
\label{WKBapprox}

The functional determinant in the one-loop term of the effective action 
(\ref{aa}) can be expressed by means of the Gelfand-Yaglom
formula \cite{gelf,klein1,klein2}
as
\begin{eqnarray}
\label{ffoonn}
&&\Gamma[q]=S[q]+\frac{i\hbar} 2  \lim_{\varepsilon\rightarrow 0^+}\lim_{T\rightarrow\infty}
\ln\left(\frac{\omega_\varepsilon F_\varepsilon(T)}
{\sin(2\omega_\varepsilon T)}\right)+{\cal O}\,(\hbar^2)
\end{eqnarray}
where $F_\varepsilon(t)$ is the solution of
\begin{eqnarray}
\label{sistdiff}
\left\{\begin{array}{l}
\ddot F_\varepsilon(t)+\left(\omega_\varepsilon^2+
\frac1m\partial_q^2U(q(t))\right)
F_\varepsilon(t)=0\\
F_\varepsilon(-T)=0\\
\dot{F}_\varepsilon (-T)=1.\\
\end{array}\right.
\end{eqnarray}
Note that the time variable appearing in the above equations is the real time.
At first sight it might seem that, on account of the factor $i$ in (\ref{ffoonn}),
the one-loop contribution to $\Gamma[q]$ is imaginary if $q(t)$ is real. Actually,
the effect of the regularization $\omega\to\omega_\varepsilon$ is such that $\Gamma_1[q]$
has generally both a real and
an imaginary part. As we shall see, the latter disappears if $q(t)$ 
varies slowly with time.
Without the regularization the expression for $\Gamma_1[q]$
would be ill-defined, both numerator and denominator oscillating
with $T$. 

We obtain the derivative expansion of the effective
action at order $\hbar$ starting from Eq. (\ref{ffoonn}).
For the moment we neglect the frequency regularization which we will
reintroduce later. 
In order to deal with convergent integrals we suppose that $q(t)=0$ for
$|t|>s$. At the end of the calculation, \idest, after the limits $T\rightarrow\infty$ and
$\varepsilon\rightarrow0^+$ have been taken, we will let 
$s\rightarrow\infty$.

If we set $q(t)=Q(\rho t)$, $q(t)$ varies
slowly if $\rho$ is small. The expansion of $\Gamma[q]$ around
$q(t)$ constant is therefore related to the asymptotic expansion of
$F(t)$ for $\rho\rightarrow 0$.
Introducing the variable $\tau=\rho t$ and setting $\Phi(\tau)\equiv
F(\tau/\rho)$ and 
$k^2(\tau) \equiv \omega^2+(1/m)\,\partial_q^2U(Q(\tau))$, 
Eq. (\ref{sistdiff}) becomes
\begin{eqnarray}
\label{wkab2}
\left\{\begin{array}{l}
\frac{d^2}{d\tau^2}\Phi(\tau)+\frac 1{\rho^2}k^2(\tau)\Phi(\tau)=0\\
\Phi(-\rho T)=0 \\
\frac{d}{d\tau}\Phi(-\rho T)=\frac1\rho.\\
\end{array}\right.
\end{eqnarray}
An approximate solution of (\ref{wkab2}) for $\rho\rightarrow 0$  can be found
by means of the WKB method \cite{froman} with the parameter $\rho$ playing the role
of $\hbar$.
The $N$-th order solution is 
\begin{equation}
\label{hhh}
\Phi_{\mbox{\tiny 2N}}(\tau)=\frac{1}{\sqrt{W_{\mbox{\tiny 2N}}(\tau)}}\left[c_+e^{\frac i\rho\int^{\tau} W_{\mbox{\tiny 2N}}(\tau')d\tau'}
+c_-e^{-\frac i\rho\int^{\tau} W_{\mbox{\tiny 2N}}(\tau')d\tau'}\right],
\end{equation}
where
$W_{\mbox{\tiny 2N}}(\tau)$ is obtained, 
neglecting all the terms of order higher than $\rho^{2N}$,
from the recursive relation 
\begin{eqnarray}
\label{hhh2}
W_{\mbox{\tiny 2N}}(\tau)=\left[k^2(\tau)+\rho^2\sqrt{W_{2(N-1)}(\tau)}
\frac{d^2}{d\tau^2}\left(\frac{1}{\sqrt{W_{2(N-1)}(\tau)}}\right)\right]^\frac12
\end{eqnarray}
with
\begin{equation}
W_0(\tau)=k(\tau).
\end{equation}
Imposing the initial conditions and going back to the variable
$t$ we find that at the lowest order the solution of
(\ref{sistdiff}) is
\begin{equation}
\label{t->T}
F_0(t)=\frac
{e^{i\int_{-T}^t dt' \sqrt{\omega^2+\frac 1m\partial_q^2U(q(t'))}}-
e^{-i\int_{-T}^t dt' \sqrt{\omega^2+\frac 1m\partial_q^2U(q(t'))}}}
{2i\sqrt{\omega \sqrt{\omega^2+\frac 1m\partial_q^2U(q(t))}}}.  
\end{equation}
When the regularization $\omega\rightarrow\omega_\varepsilon$ is 
reintroduced
and $F_0(t)$ is evaluated at the  time $T$,
the second exponential, proportional to $e^{-2i\omega_\varepsilon(T-s)}$,
vanishes for large $T$ and can be neglected, since the limit 
$T\rightarrow\infty$
has to be performed before the limit $\varepsilon\rightarrow0^+$. 
We obtain therefore
\begin{equation}
\label{pote}
\Gamma[q]\simeq S[q]-\frac\hbar 2\int_{-s}^s dt \left(\sqrt{\omega^2+
\frac1m\partial_q^2U(q(t))}-\omega\right).
\end{equation}
Recalling (\ref{qpote}), the first quantum correction to the classical
potential is 
\begin{equation}
\label{pote5}
V_{e1}(q)=\frac 1 2\left(\sqrt{\omega^2+\frac1m\partial_q^2U(q)}-\omega\right).
\end{equation}
The next order of the WKB approximation gives
\begin{eqnarray}
\label{zetaca}
\Gamma[q]&\simeq&S[q]-\frac\hbar 2\int_{-s}^s dt \left(\sqrt{\omega^2+
\frac1m\partial_q^2U(q(t))}-\omega\right) \\ 
&&+\frac\hbar 2\int_{-s}^s dt
\frac 1 {32m^2}\frac{(\partial_q^3U(q(t)))^2}{\left(\omega^2+\frac1m\partial_q^2U(q(t))\right)^\frac 52}\dot q^2(t)
\nonumber
\end{eqnarray}
which implies
\begin{equation}
\label{zetaca5}
Z_1(q)=\frac 1 {32m^2}\frac{(\partial_q^3U(q))^2}
{\left(\omega^2+\frac1m\partial_q^2U(q)\right)^\frac 52}.
\end{equation}
Equations (\ref{pote5}) and (\ref{zetaca5}) agree with the results 
found in Section \ref{aiut!}.

If the classical potential $V(q)$ is not everywhere convex,
in the regions where $\omega^2+\frac 1m\partial_q^2U(q)$ is negative
the effective potential $V_{e1}(q)$ and $Z_1(q)$ become imaginary.
Moreover, $Z_1(q)$ has a divergence at the points where 
where $\omega^2+\frac 1m\partial_q^2U(q)=0$ and this corresponds 
to the fact that the WKB approximation loses its
validity near the turning points $k^2(\tau)=0$.

It is clear that the $N$-th order WKB approximation for
$\Phi(\tau)$ corresponds to the derivative expansion of $\Gamma_1[q]$ 
at order $2N$.
One can also check that no terms with an odd number of
derivatives appear.
The connection to the WKB approximation also shows, as previously stated,
that the derivative expansion has only an asymptotic validity for $\dot q\to0$.

From Eq. (\ref{ffoonn}) it
is clear that if $q(t)$ is real, $\Gamma[q]$ up to one-loop order 
is not necessarily real.
However, from Eq. (\ref{hhh2}) we see that, if the classical
potential is everywhere convex, all the terms of the derivative expansion
 of the effective action are real if $q(t)$ is real. 
The contradiction is only apparent. 
It can be seen that the imaginary part of $\Gamma[q]$ is due to
singularities in the Green functions which do not contribute to
the derivative expansion.

In conclusion, in the case of vanishing external source $J(t)=0$
the DE2 approximation at order $\hbar$ of Eq. (\ref{eulaq})
reads
\begin{eqnarray}
\label{cauchi}
\left(m+\hbar Z_1(q(t))\right)\ddot q(t)+\frac{\hbar}{2} 
\partial_q Z_1(q(t)) \dot q^2(t)=
-\partial_q \left(V(q(t))+\hbar V_{e1}(q(t))\right).
\end{eqnarray}
We discuss the validity of this equation in the case 
$V(q)=\frac{1}{2} m\omega^2q^2+\frac g{4!}q^4$.
Equation (\ref{cauchi}) is approximate both because the DE2 approximation
is adopted and because the terms of order higher than $\hbar$ are 
neglected.
For a solution $q(t)$
of amplitude $A$ these two approximations are valid if 
\begin{equation}
\label{stima}
\frac{\frac g{4!}A^4}{\frac{1}{2} m\omega^2A^2}\ll 1
\end{equation}
and
\begin{equation}\label{stima2}
\frac{\hbar g}{m^2\omega^3}\ll1,
\end{equation}
respectively.
Under these conditions the solutions of Eqs. (\ref{cauchi}) and
(\ref{eulaq}) remain close for a time $t$ satisfying 
\begin{equation}
\label{stima212}
\omega t ~ \frac{\hbar g}{m^2\omega^3}~
\frac{\frac g{4!}A^4}{\frac{1}{2} m\omega^2A^2}\ll1.
\end{equation}

\section*{PART II}
\section{2-D anharmonic oscillator: classical}
\label{2-D}

Classical systems with more than one degree of freedom present 
a richer variety of phenomena and in particular they may exhibit
chaotic behavior for $J=0$.
The formalism described in Part I can be generalized without difficulties
to many degrees of freedom.
Here we study the system whose Lagrangian is
 \cite{pe}
\begin{equation}
L(\dot{q}_1,\dot{q}_2,q_1,q_2)=\frac{1}{2}m(\dot{q}_1^2+\dot{q}_2^2)-
\frac{1}{2}m\omega^2(q_1^2+q_2^2)-gq_1^2q_2^2.
\label{cl}
\end{equation}
Apparently, the system has four free parameters: $m$, $\omega$, $g$ and 
the energy $E$.
However, the rescaling $t\rightarrow t/\omega,q_i\to q_i
\sqrt{m\omega^2/g},\dot{q}_i\rightarrow \dot{q}_i\sqrt{m\omega^4/g},$ for $i=1,2$,  yields
\begin{equation}
L \rightarrow 
\frac{m^2\omega^4}{g}\left[\frac{1}{2}(\dot{q}_1^2+\dot{q}_2^2)-
\frac{1}{2}(q_1^2+q_2^2)-q_1^2q_2^2\right],
\label{lscalata}
\end{equation}
where, now, $\dot{q}_i$, $q_i$ and $t$ are dimensionless.
The energy of the system (\ref{cl}) is then $E=(m^2\omega^4/g)\varepsilon$, where $\varepsilon$ is the dimensionless energy of the dimensionless 
Lagrangian $L=\frac{1}{2}(\dot{q}_1^2+\dot{q}_2^2)-
\frac{1}{2}(q_1^2+q_2^2)-q_1^2q_2^2$.
We conclude that $\varepsilon$ is the unique free parameter of the system 
under consideration.

The rescaled equations of motion 
\begin{eqnarray}
\ddot{q}_1 &=& - q_1 (1+2q_2^2) \\
\ddot{q}_2 &=& - q_2 (1+2q_1^2)
\end{eqnarray} 
have been numerically integrated using a standard fourth order 
Runge-Kutta method \cite{numrec}.
A qualitative description of the corresponding solutions has been
achieved by constructing the surfaces of section (Poincar\'e sections)
\cite{Henon} and evaluating the largest Lyapunov exponent \cite{notalyap}.
The degree of chaoticity of the system can be summarized by the fraction
of regular orbits on the energy shell as a function of the dimensionless 
energy $\varepsilon$.
This fraction is close to unity for $\varepsilon \lesssim 0.75$
and vanishes exponentially for $\varepsilon \gtrsim 0.75$.
The border value $\varepsilon= 0.75$ agrees with that obtained
from the Toda criterion \cite{toda}.
In our system, the sign of the curvature of the energy surface where 
the motion takes place is given by
$\mbox{sign}(\det\mbox{He}(V))$, 
where He($V$) is the Hessian of $V=(q_1^2+q_2^2)/2+q_1^2 q_2^2$, \idest,
$\mbox{He}(V)_{ij}=\frac{\partial^2 V}{\partial q_i\partial q_j}$.
This sign changes from positive to negative at $\varepsilon=3/4$.
It is worth noting that the Toda criterion does not detect the 
first occurrence of chaos in the Poincar\'e sections 
\cite{notacontrotoda,casati}. 
Nevertheless, when $\varepsilon > 3/4$ 
we find that chaotic orbits are spread all over the sections.
For $\varepsilon < 3/4$, the irregular orbits are located 
in a small region of the Poincar\'e sections, namely near a perturbed 
separatrix where chaos initially appears in consequence of the mechanism 
of the heteroclinic intersection \cite{lieberman}.

\section{2-D anharmonic oscillator: quantum}
\label{2-D2}

In the semiclassical and local approximations,
the quantum system corresponding to the 2-D anharmonic oscillator
introduced in the previous Section is described by an effective
Lagrangian (effective action density) 
\begin{eqnarray}\label{le}
L_e(\dot{\bf q},{\bf q})=\frac{1}{2}Z_{ij}({\bf q})
\dot{q}_i\dot{q}_j -V_e({\bf q}),
\end{eqnarray} 
where
${\bf q}=(q_1,q_2)$ and 
$\dot{\bf q}=(\dot{q}_1,\dot{q}_2)$.
In the rescaled variables used in the classical case, we have
\begin{eqnarray}
&&V_e=\frac{1}{2} (q_1^2+q_2^2)+q_1^2q_2^2+
\frac{\gamma}{2} \left(\sqrt{\Lambda_+}+\sqrt{\Lambda_-}
-2\right) \label{refpo}\\
&&Z_{11}=1+\gamma \left\{\frac{q_1^2}{8}\left[\frac{(1+\eta)^2}
{\Lambda_+^{5/2}}+\frac{(1-\eta)^2}
{\Lambda_-^{5/2}}\right]+8q_2^2\zeta\right\}
\\
&&Z_{12}=Z_{21}=
\gamma \left\{\frac{q_1q_2}{8}\left[\frac{(1+\eta)(1+\xi)}
{\Lambda_+^{5/2}}+\frac{(1-\eta)(1-\xi)}
{\Lambda_-^{5/2}}\right]+8q_1q_2\zeta\right\}
\\
&&Z_{22}=1+\gamma \left\{\frac{q_2^2}{8}\left[\frac{(1+\xi)^2}
{\Lambda_+^{5/2}}+\frac{(1-\xi)^2}
{\Lambda_-^{5/2}}\right]+8q_1^2\zeta\right\},
\end{eqnarray}
where 
\begin{eqnarray}
\Lambda_{\pm} = 1+q_1^2+q_2^2\pm\Sigma,
\end{eqnarray}
\begin{eqnarray}
\Sigma = \sqrt{q_1^4+q_2^4+14q_1^2q_2^2}, 
\end{eqnarray}
\begin{eqnarray}\label{eta}
\eta = (q_1^2+7q_2^2)/\Sigma,
\end{eqnarray}
\begin{eqnarray}\label{xi}
\xi = (q_2^2+7q_1^2)/\Sigma,
\end{eqnarray}
and
\begin{eqnarray}\label{zeta}
\zeta = {[( q_1^2+q_2^2)/ \Sigma]^2 \over
[ \sqrt{\Lambda_+\Lambda_-} (\sqrt{\Lambda_+}+
\sqrt{\Lambda_-})^3]}.
\end{eqnarray}
With respect to the classical system, we have an additional  parameter 
$\gamma=\hbar g/ m^2\omega^3$ which arises from rescaling $\hbar$.
It can be seen that, when ${\bf q}$ varies, the 
effective potential and the symmetric kinetic matrix $Z_{ij}$ 
can be singular or complex-valued, unless ${\bf q}$ is constrained 
inside a certain region. 
If we limit ourselves to the region where the effective 
potential, in the considered approximation, is convex, then $L_e$ is
well defined.

\begin{figure}[!h]
\begin{center}
\includegraphics*[width=10cm]{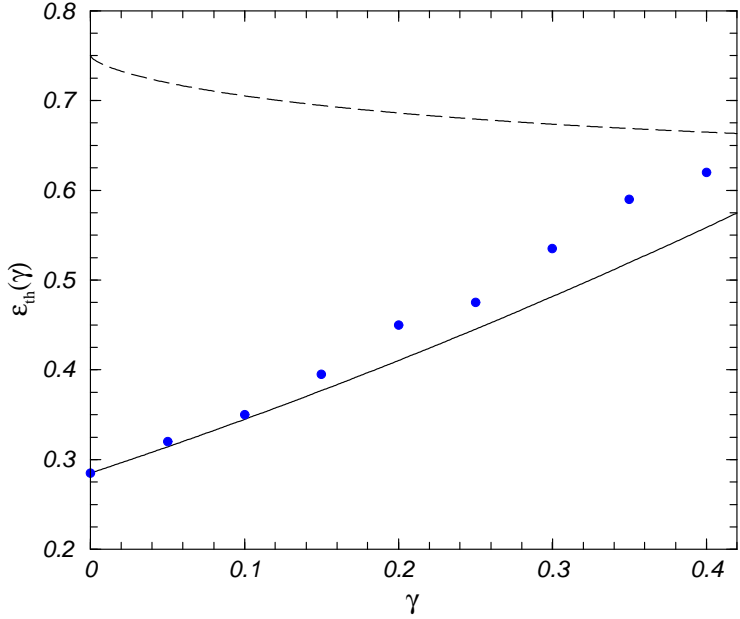}
\end{center}
\caption{Threshold energy $\varepsilon_{th}$ at which chaotic behavior
shows up in the Poincar\'e sections of the effective system 
{\it versus} $\gamma = \hbar g /m^2 \omega^3$. 
The solid line is the theoretical estimate 
$\varepsilon_{th}(\gamma)=\varepsilon_{th}(0)(1+\gamma)^2$.
The dashed line represents the maximal energy $\varepsilon_m$ below which 
the effective potential is everywhere convex.}
\protect\label{emaxg}
\end{figure}

The rescaled equations of motion corresponding to the Lagrangian 
(\ref{le})
\begin{eqnarray}
~~~~~~~~~~Z_{11} \ddot{q}_1 + Z_{12}\ddot{q}_2 =
  -\frac{1}{2} \frac{\partial Z_{11}}{\partial q_1}
  \dot{q}_1 \dot{q}_1 - \frac{\partial Z_{11}}{\partial q_2}
  \dot{q}_1 \dot{q}_2 + \left( \frac{1}{2}\frac{\partial Z_{22}}
    {\partial q_1} - \frac{\partial Z_{12}}{\partial q_2}  \right)
  \dot{q}_2 \dot{q}_2
  -\frac{\partial V_e}{\partial q_1} 
\end{eqnarray}
\begin{eqnarray}
~~~~~~~~~~
Z_{12} \ddot{q}_1 + Z_{22}\ddot{q}_2 =
  -\frac{1}{2} \frac{\partial Z_{22}}{\partial q_2}
  \dot{q}_2 \dot{q}_2 - \frac{\partial Z_{22}}{\partial q_1}
  \dot{q}_1 \dot{q}_2 + \left( \frac{1}{2}\frac{\partial Z_{11}}
    {\partial q_2} - \frac{\partial Z_{12}}{\partial q_1} \right)
  \dot{q}_1 \dot{q}_1
  -\frac{\partial V_e}{\partial q_2} 
\end{eqnarray}
have been numerically solved as in the classical case.
These equations are nonlinear and may lead to a chaotic evolution.
However, due to the fact that no chaotic behavior is allowed at quantum 
level, we expect a reduction of chaoticity in the effective system
with respect to the classical one. 
This reduction should depend on the value of the parameter $\gamma$, the 
value $\gamma=0$ corresponding to the classical system.
In Fig. \ref{emaxg} we illustrate, for different values of $\gamma$, 
the smallest energy (threshold energy) $\varepsilon_{th}$ at which chaos 
shows up in the Poincar\'e sections of the effective system 
\cite{notasoglia}.
We see that $\varepsilon_{th}$ increases with increasing $\gamma$.
This behavior can be explained as follows.
Let us consider the Taylor expansion of the effective Lagrangian 
around ${\bf q}={\bf 0}$ and $\dot{\bf q}={\bf 0}$.
Up to quadratic terms, we obtain 
$L_e(\dot{\bf q},{\bf q})=\frac{1}{2}(\dot{q}_1^2+\dot{q}_2^2)
- \frac{1}{2}(1+\gamma)(q_1^2+q_2^2)+ \ldots$.
In terms of unrescaled variables this corresponds to a shift
of the classical frequency $\omega \to \omega \sqrt{1+\gamma}$.
The rescaled energy $\varepsilon=E/(m^2\omega^4/g)$ picks up a
factor $(1+\gamma)^2$. 
This means that if $\varepsilon_{th}(0)$ denotes the threshold energy 
at $\gamma=0$, we should have approximately 
$\varepsilon_{th}(\gamma) = \varepsilon_{th}(0)(1+\gamma)^2$.
This prediction is well confirmed in Fig. \ref{emaxg}.
It is easy to see that the increase of the threshold for chaos holds 
under the general condition that $V_{e1}$ is convex, which, in turn, 
amounts to 
\begin{equation}
\mbox{He}\left(\mbox{tr}
\left(\mbox{He}(V)\right)^{\frac{1}{2}}\right) > 0.
\end{equation}
where $V$ is the classical potential.
These results parallel those obtained in \cite{cgm} for the 
$N$-component $\phi^4$ oscillators where the mean field plays 
the role of classical system.
 
The range of $\gamma$ values explored in Fig. 1 includes situations
encountered in molecular physics.  
In fact, the vibrational Hamiltonian of diatomic molecules is often
assumed as a quartic oscillator and using
the numerical values of Ref. \cite{WMT} obtained from spectroscopic
data we find that $10^{-4} \lesssim \gamma \lesssim 10^{-1}$.

In the following we compare the solutions of the local effective 
equations with the classical solutions and with the exact quantum
evolutions of coherent states centered at the initial conditions
of the local equations.
We have already remarked that the initial conditions for the classical
and the effective dynamics do not determine completely the initial 
wave function but provide only a constraint.
Therefore the choice of the initial wave function is not unique.
A natural choice is represented by a harmonic coherent state which
is parametrized by the expectation value of position and
momentum. 
In fact, by performing simulations with initial wave functions 
which satisfy the proper constraints but are of arbitrary shape
we find that the agreement between the effective and quantum dynamics is 
very poor when the shape of the initial wave function differs
substantially from that of a coherent state.

In rescaled units, the exact quantum dynamics is defined by the 
Schr\"odinger equation
\begin{equation}
i \gamma \frac{\partial}{\partial t} \psi(q_1,q_2,t) = 
\hat{H} \psi(q_1,q_2,t),
\label{quantumevol}
\end{equation}
with
\begin{eqnarray}
\hat{H} &=& 
\frac{1}{2}(\hat{p}_1^2+\hat{p}_2^2)+
\frac{1}{2}(\hat{q}_1^2+\hat{q}_2^2)+\hat{q}_1^2\hat{q}_2^2,
\label{hqres}
\end{eqnarray}
where $\hat{p}_j=-i\gamma\frac{\partial}{\partial q_j}$ and 
$\hat{q}_j=q_j$, for $j=1,2$, are the rescaled momentum and 
position operators. 
In order to solve (\ref{quantumevol})  we represent the rescaled 
Hamiltonian operator (\ref{hqres})
in the basis of the eigenstates of the associated 2-D harmonic 
oscillator $\hat{H}_0 = \frac{1}{2}(\hat{p}_1^2+\hat{p}_2^2)+
\frac{1}{2}(\hat{q}_1^2+\hat{q}_2^2)$.
The corresponding infinite matrix is truncated and then diagonalized 
with standard techniques \cite{numrec}. 
As initial state we choose the coherent state 
\begin{equation}
|p_1'q_1'p_2'q_2' \rangle = 
e^{-{i\over\gamma} q_1' \hat{p}_1} 
e^{ {i\over\gamma} p_1' \hat{q}_1} 
e^{-{i\over\gamma} q_2' \hat{p}_2} 
e^{ {i\over\gamma} p_2' \hat{q}_2} | 0_0 \rangle, 
\end{equation}
where $| 0_0 \rangle$ is the ground state of $\hat{H}_0$.
The parameters $p_1'q_1'p_2'q_2'$ are taken equal to the initial 
conditions used in the integration of the classical and effective 
Lagrangians.
\begin{figure}
\begin{center}
\centerline{\includegraphics*[width=13.5cm]{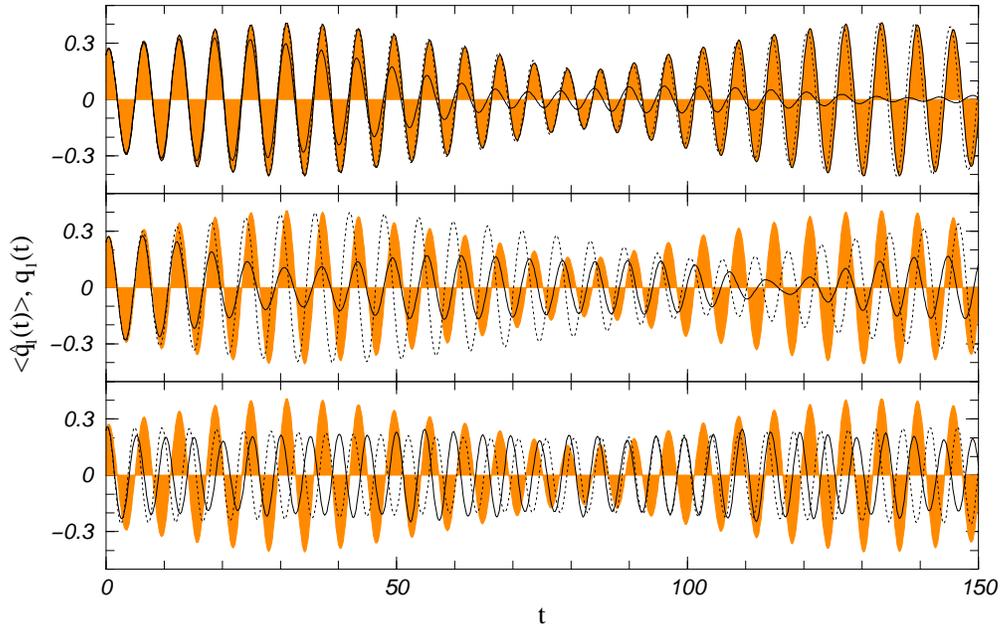}}
\end{center}
\caption{Time evolution of the expectation value of the 
position operator $\langle \hat{q}_1(t) \rangle$ (solid line) 
compared with the classical (shaded area) and the effective (dots)
solutions $q_1(t)$.
The three panels correspond, from top to bottom, to 
$\gamma=0.01$, 0.1, and 1, respectively.
In all cases we have a classical rescaled energy $\varepsilon=0.1$.}
\protect\label{timeconfronto}
\end{figure}

In Fig. 2 we show the evolution of $\langle \hat{q}_1(t) \rangle$ 
in comparison with the corresponding classical and effective solutions.
The three panels correspond, from top to bottom,  
to increasing values of $\gamma$ at constant dimensionless 
classical energies $\varepsilon$.
Note that the initial coherent state depends on $\gamma$.

Figure 2 shows that by increasing $\gamma$ there is a crossover 
in the behavior of the solution of the effective dynamics.
At small $\gamma$ the effective solution stays close to the 
classical one while for larger $\gamma$ it reproduces qualitatively 
the shape of the quantum evolution.
We notice that in the large-$\gamma$ region the quantum and the
effective dynamics do not show, on the time scale considered,
a transfer of energy among the degrees of freedom as the 
classical solution.
This seems to indicate that the quantum corrections in the effective 
dynamics have an anti-mixing influence that regularizes the motion.
Of course, over longer times a transfer of energy
takes place also in the quantum and effective evolutions.
The theoretical implications of these results will be discussed 
elsewhere.


\begin{thebibliography}{99}
\bibitem{notalyap} In a conservative system with two degrees of freedom, 
only the largest exponent matters, for two exponents are always zero 
and the other two are one the opposite of the other.
\bibitem{notasoglia} The definition of a threshold energy is clearly not
unique. We systematically inspected the Poincar\'e sections $q_2=0$
for different values of $\gamma$ and energy. 
Since chaos shows up at $p_1 \simeq 0$, we considered the ensemble
of trajectories obtained with initial conditions $p_1(0)=0$, 
$p_2(0)$ determined by energy conservation and 
$q_1(0)$ varied with a step $5 \times 10^{-3}$.
For each value of $\gamma$, we explored the energy range where the
effective potential is everywhere convex
with a step $5 \times 10^{-3}$.
\end{thebibliography}
\end{document}